\documentclass[a4paper, amsfonts, amssymb, amsmath, reprint, showkeys, twoside,superscript address]{revtex4-2}
\usepackage{graphicx} % Required for inserting images
\usepackage{bm,amssymb,amsmath}
\usepackage{graphicx,color,upgreek}
\usepackage{float}
\usepackage{xcolor}
\usepackage{relsize}
\begin{document}
\title{Efficient  protocol   for the Markovian Mpemba effect in $N$-level systems} 
\author{David Gelbwaser-Klimovsky}
 \email{\tt dgelbi@technion.ac.il}
 \affiliation{Schulich Faculty of Chemistry and Helen Diller Quantum Center, Technion-Israel Institute of Technology, Haifa 3200003, Israel} \begin{abstract}
 	The Mpemba effect is an anomalous phenomenon in which the time to thermalize does not depend monotonically on the distance from equilibrium, allowing systems with an initially larger temperature difference to thermalize before those with a smaller one. The rarity of the effect makes it hard to find parameters that produce it, complicating the design of experiments and the development of applications. Here, we find necessary and sufficient conditions for the Markovian Mpemba effect in three-level systems and explain its underlying physical mechanisms.  Based on our understanding of the three-level case, we develop an efficient algorithm to determine the parameters for the effect in  $N$-level systems. This protocol could open the door for the realization of Mpemba experiments and related applications in large systems.
 \end{abstract}
\maketitle

The Mpemba effect is a thermodynamic anomaly in which a system farther from thermal equilibrium thermalizes faster than one initially closer to it. The best-known example of the effect is in water, where a higher initial temperature sometimes results in a shorter freezing time \cite{mpemba1969cool}.
Besides water, other systems exhibit this anomalous thermalization. This includes a wide range of systems, such as trapped ions \cite{aharony2024inverse},  granular gases \cite{biswas2020mpemba}, spin glasses \cite{baity2019mpemba}, and colloids \cite{kumar2020exponentially}. There is a recent surge of interest in the subject of anomalous or accelerated thermalization \cite{ivander2023hyperacceleration,moroder2024thermodynamics,strachan2025non,bera2023effect,degunther2022anomalous,biswas2025mpemba,blum2025thermalization,chatterjee2024multiple,klich2019mpemba,ramon2025thermal,lu2017nonequilibrium}. An important part of the knowledge in the area has being reviewed in \cite{teza2026speedups,ares2025quantum}. Furthermore, multiple applications of the Mpemba effect have been proposed, such as improving heating/cooling protocols \cite{gal2020precooling}, minimizing dissipation \cite{walker2023optimal}, boosting sensors' performance \cite{pagare2024mpemba}, slowing biochemical reactions \cite{hatakeyama2024enzymatic}, and improving Monte Carlo algorithms \cite{klich2018solution}.

A key challenge in the field is determining which systems, and under which parameter regimes, exhibit the Mpemba effect. This is necessary to understand the relevant physical mechanisms better, conduct experiments and develop applications. In particular, for Markovian quantum or classical discrete systems, a Pauli rate equation describes thermalization. Although the parameter space of this equation can be explored for a few levels, the problem becomes intractable for large systems. To address this, machine learning has been used, but so far it has been restricted to the Ising model \cite{amorim2023predicting}.
Recently, another approach has been used. In \cite{avitan2026necessary}, the necessary conditions for the Markovian Mpemba effect in three-level systems (3LSs) were derived. Moreover, it was shown that violating these conditions can be used to rule out the Markovian Mpemba effect in $N$-level systems and to explain why the effect is a thermodynamic anomaly. Nevertheless, that work stopped short of finding which $N$-level systems exhibit the Markovian Mpemba effect.

In this work, we revisit the  3LS and derive the if-and-only-if conditions for the Markovian Mpemba effect. This helps us study the physical mechanisms at play in 3LSs. In particular, we find that every 3LS has a combination of parameters, which we call a singular point, where the thermalization is governed by a single non-zero thermalization speed rather than two. The 3LS singular point is partially surrounded by a parameter region that exhibits the Markovian Mpemba effect. Therefore, a small deviation of parameters from the singular point in the correct direction produces the effect. Inspired by this, we analytically identify the singular point for any $N$-level system and use it to efficiently find parameters for the effect by randomly choosing a single pair of states and varying the corresponding forward and backward rates. This protocol could guide the design of Mpemba experiments or applications in large systems.

\section{Results} To analyze the Markovian Mpemba effect in $N$-level systems, we start by studying the smallest example that can present the effect: a three-level system. We first identify the basic Mpemba principles and then examine whether they extend to larger systems. This approach has already been shown to be effective for studying some properties of the Markovian Mpemba effect in $N$-level systems \cite{avitan2026necessary}.  

The basic setup is the following: a non-degenerate 3LS interacting with a thermal bath at inverse temperature $\beta_b$. The 3LS energy levels are $\epsilon_3>\epsilon_2>\epsilon_1$. Under standard open quantum systems and Markovian approximations \cite{breuer_theory_2002}, the Pauli rate equation governs the thermalization dynamics $\mathbf{\dot{P}}=M\mathbf{P}$, where $\mathbf{P}$ is the population vector and $M$ is the transition matrix. Its matrix elements  are $M_{ii}=-\sum_j a_{ji}$ and $M_{i\neq j}=a_{ij}$. Here, $a_{ij}$ is the transition rate from state $j$ to state $i$, and we assume it satisfies detailed balance, although this is not strictly necessary for thermalization \cite{alicki2023violation,blum2025thermalization}. We call $\lambda_i$ the transition matrix eigenvalues, which are ordered as $0=\lambda_1>\lambda_2\geq \lambda_3$, and we denote the respective eigenvectors as $\mathbf{v_i}$. Note that, in addition to non-degenerate open quantum systems, Pauli rate equations can also describe thermalization in classical discrete systems. Therefore, our results also apply to those systems.

\begin{figure}[htbp]
		\centering
		\includegraphics[width=1\linewidth]{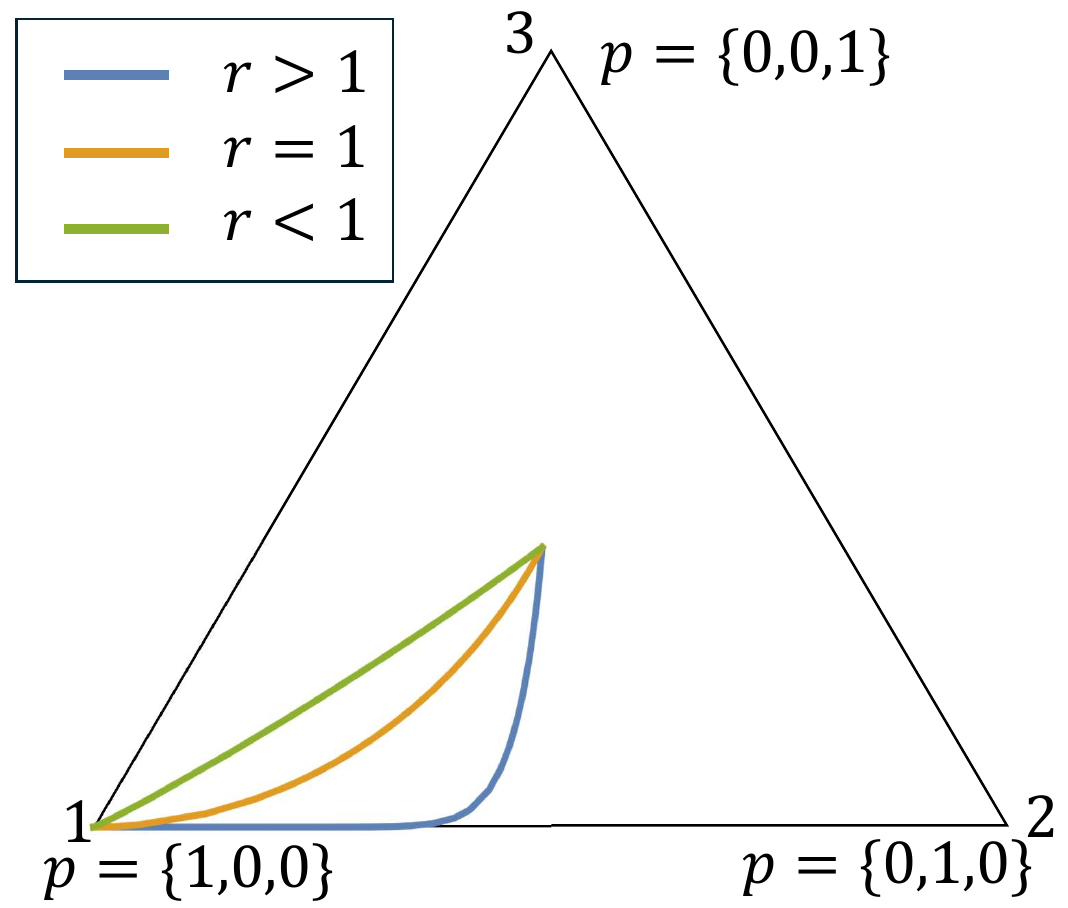}
		\caption{Each point inside the triangle represents a state of the 3LS population, $\{p_1,p_2,p_3\}$. The corners are pure states of levels 1, 2 or 3, respectively. The thermal states are represented by one of the color lines, whose shape depends on the energy level distribution through $r$.}	\label{fig:tri}
	\end{figure}

The 3LS population states can be represented in a 2D equilateral triangle (see Figure \ref{fig:tri}).  For simplicity, we rotate the population states into the $x-y$ plane and associate the origin with the infinite-population state (see supporting information). In this space, all thermal states lie on a line known as the quasistatic locus. For any distance definition that always contracts with time, increases with initial temperature,  is continuous and convex, the 3LS exhibits the Markovian Mpemba effect if and only if $\lambda_2\neq\lambda_3$ and the fast eigenvector, $\mathbf{v_3}$, is parallel to a tangent of the quasistatic locus (see  \cite{lu2017nonequilibrium} and supporting information). The angle of the quasistatic locus tangent relative to the triangle base ranges from $0$ to $ \arctan\left(\frac{1+2r}{\sqrt{3}} \right) $, where $r=\frac{\epsilon_3-\epsilon_2}{\epsilon_2-\epsilon_1}$. Therefore, the iff condition for the Markovian  Mpemba effect is 

\begin{equation}
     0< \frac{(\mathbf{v_3})_y}{(\mathbf{v_3})_x}<\frac{1+2r}{\sqrt{3}},
\end{equation}
where $(\mathbf{v_3})_i$ is the $i$ component of the fast eigenvector in the $x-y$ plane. Using the analytical expression for the fast eigenvector (see supporting information), we can rewrite the condition in terms of the transition rates as

\begin{equation}
 -2r_f<l+Sign[p]\sqrt{l^2+q} ,\label{eq:cond}
\end{equation}
where $p=e^{-\beta_b(\epsilon_3-\epsilon_2)}(a_{13}e^{-\beta_b(\epsilon_2-\epsilon_1)}-a_{23})$, $ q=4 (a_{12}e^{-\beta_b(\epsilon_2-\epsilon_1)} - a_{23})/p$ and $ l=(a_{12}-a_{13})/p+ q/4-1$ and  $r_f=\frac{r-1}{1+2r}$. Eq. \ref{eq:cond} shows that as $r$ increases, the Markovian Mpemba conditions are less restrictive. This should not come as a surprise because the maximal angle of the quasistatic locus increases with  $r$. As we show, $r=1$, which corresponds to equal energy-level spacing, defines a boundary between distinct physical mechanisms.  

From Eq. \eqref{eq:cond} conditions on $p$, $q$ and $l$ for the Markovian Mpemba effect can be derived. These quantities have a specific dependence on the transition rates, and
not all the mathematically valid conditions on $p$, $q$ and $l$ can be satisfied once the explicit dependences on the transition rates and detailed balance are taken into account (e.g., if $p<0$ and $q<0$ then $l$ can not be positive). Therefore, there are four relevant cases in which Eq. \eqref{eq:cond} holds. Each of them represents a different physical mechanism that produces the Markovian Mpemba effect and holds for different physical parameters. The four Markovian Mpemba mechanisms are:
\begin{enumerate}
    \item   $r\geq 1$, $p>0$ and $q>0$;
    \item  $r>1$, $p>0$, $q<0$  and  $l<-r_f+\frac{q}{4r_f}$;
    \item  $r>1$, $p<0$, $q>0$  and  $l>-r_f+\frac{q}{4r_f}$;
    \item $r<1$, $p>0$, $q>-4 r_f^2$ and $l>-r_f+\frac{q}{4r_f}$.
\end{enumerate}

Notice there is a common feature among Mechanisms 1, 2 and 4. All of them require $p$ to be positive. This requirement has important physical implications. At low temperatures a positive $p$ requires $a_{13} / a_{23}\gg1$. One can argue that, for any real physical system, transition rates are bounded by some constraint and cannot take any arbitrary value. If, in addition, we consider that $a_{23}\neq0$, then our conditions imply that Mechanisms 1, 2 and 4 do not exist at low temperatures, $\beta_b(\epsilon_2-\epsilon_1)\gg1$. In this regime, only Mechanism 3 applies, which requires $r>1$. Therefore, under the constraints mentioned above,  3LSs  with $r<1$, such as the lowest 3 levels of the hydrogen atom, do not exhibit the Markovian Mpemba effect.

\begin{figure}[htbp]
		\centering
		\includegraphics[width=1\linewidth]{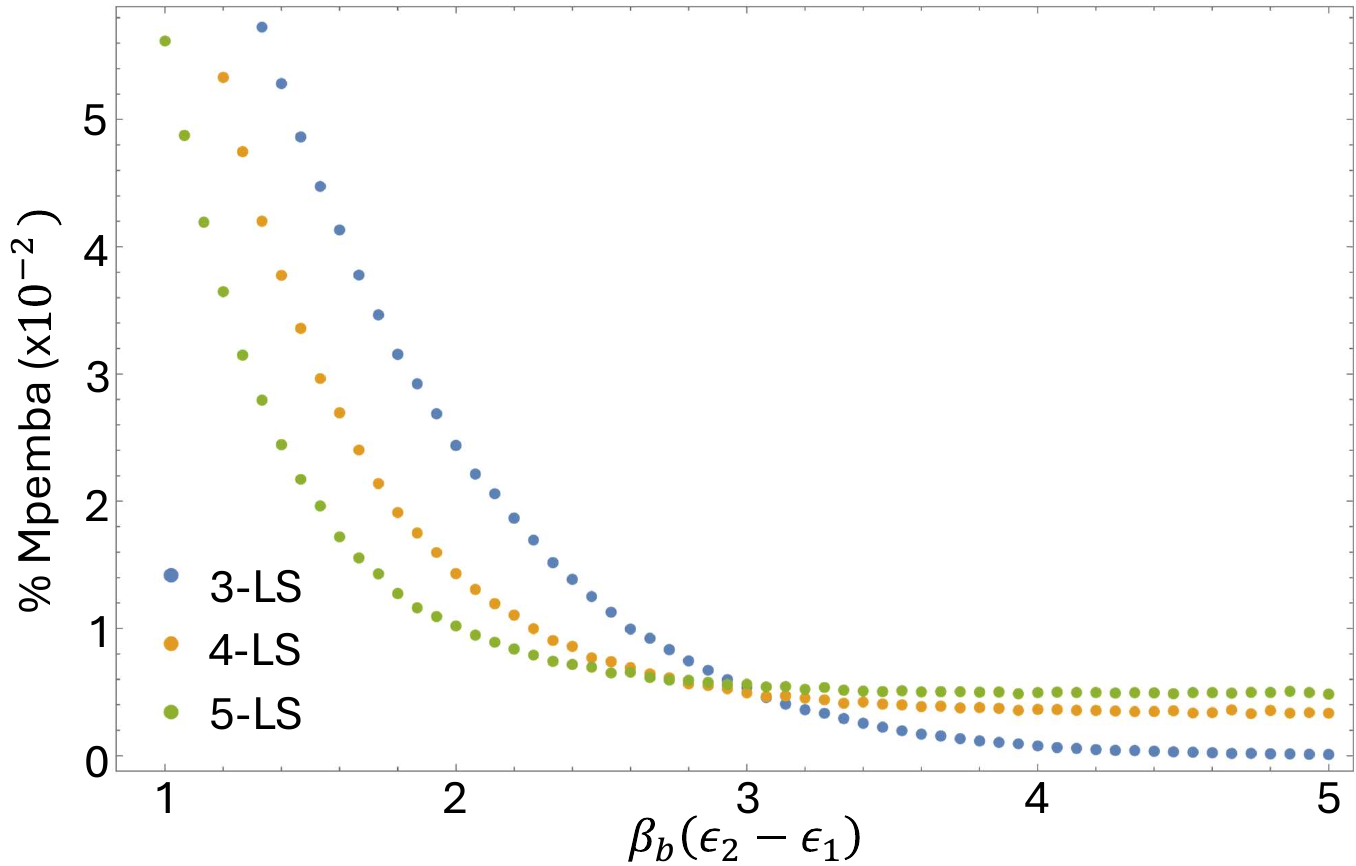}
		\caption{Probability for $ N$-level Markovian Mpemba effect as a function of the normalized bath inverse temperature. We considered bounded values of transition rates ($0<a_{ij}<1$). As shown in the plot, the probability decreases with bath temperature. $3$-, $4$-, and $ 5$-level systems are shown in the  figure. The convergence of the results was assessed by increasing the number of realizations and confirming that this did not significantly alter the plot.  For the transition rates, we considered values from 0.001 to 1 and for $\beta_b \epsilon_i$ from $0$ to $5N$. Each point in the graphs is based in $10^6$ random realizations. }	\label{fig:lowt}
	\end{figure} 
Inspired by the reduction of the active mechanism with temperature, we study how prevalent the Mpemba effect is at low temperatures when the transition rates are bounded to a specific value range. We found that the probability of the Markovian Mpemba effect decreases as temperature decreases. 
This result is not limited to 3LSs. Our numerical calculations indicate that it also holds for larger systems, such as 4- and 5-level systems. These calculations were done as follows. We started by choosing random values for energy levels ($\epsilon_1<...<\epsilon_n$) and transition rates. For the transition rates, we required them to be non-negative, satisfy detailed balance, and bound their maximum value to 1. As shown in Figure \ref{fig:lowt}, the probability of Markovian Mpemba decreases as temperature decreases. 

Next, we analyze the dependence of the mechanisms on the energy-level distribution, which is parametrized by $r$. This parameter depends on the nature of the physical systems. For example, consider only the lowest three states of the following energy level distributions: for rotational  $r>1$; for vibrational  $r=1$;  and for a hydrogen-like atom $r<1$. As mentioned above, large $r$ values are favorable for the Markovian Mpemba effect. This is reflected in the fact that three mechanisms exist for systems with $r>1$ and only one for systems with $r\leq1$.  

\begin{figure}[htbp]
		\centering
		\includegraphics[width=1\linewidth]{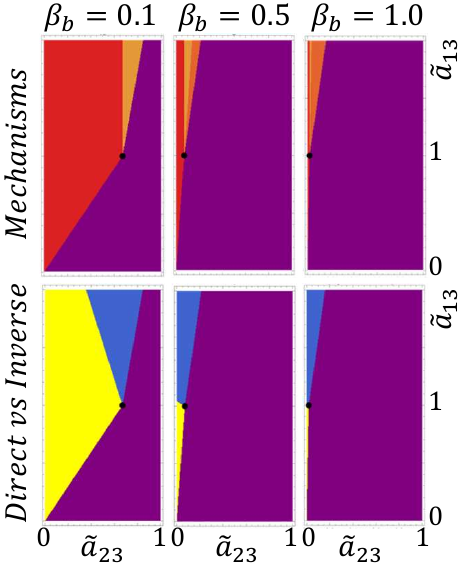}
		\caption{Markovian Mpemba effect for the lowest three rotational energy levels, $r>1$, as a function of the normalized transition rates.  The upper plots show the different mechanisms: Mechanism 1 (red), Mechanism 2 (light orange), and Mechanism 3 (Dark orange). The lower plots show the regions of the inverse (yellow) and direct (blue) Markovian Mpemba effect. In all plots, the purple regions correspond to parameter values without the Markovian Mpemba effect, and a black point represents the singular point. Each plot column corresponds to a different $\beta_b$. The energy levels are 2, 6 and 12 (in arbitrary units).}	\label{fig:rot}
	\end{figure}

\begin{figure}[htbp]
		\centering
		\includegraphics[width=1\linewidth]{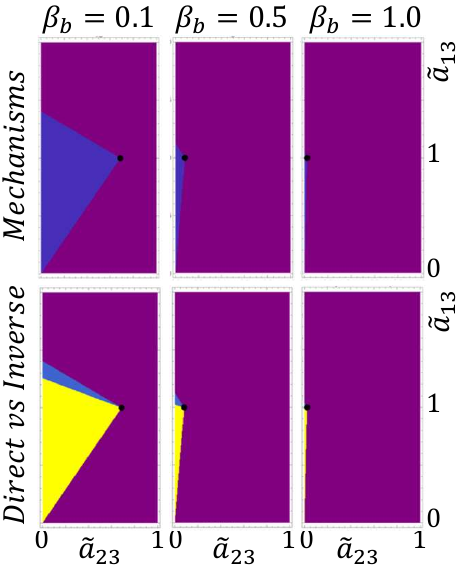}
		\caption{Markovian Mpemba effect for the lowest three hydrogen-like energy levels, $r<1$, as a function of the normalized transition rates.  The upper plots show the area (in dark blue) of Mechanism 4, the only mechanism for the Markovian Mpemba effect for $r<1$. The bottom plots show the region of the inverse Markovian Mpemba effect (yellow) and the direct Markovian Mpemba effect (blue). Each plot column corresponds to a different $\beta_b$. See Figure \ref{fig:rot} for other details of the color coding. The energy levels are (arbitrary units): -5,-5/4,-5/9. }	\label{fig:hyd}
	\end{figure}

To better understand the physics of each mechanism, we examine the transition rates and temperatures that give rise to it. We start by renormalizing the transition matrix by dividing it by the transition rate $a_{12}$. This step does not change the eigenvectors' direction. Therefore, once we assume detailed balance and fix the energy levels and temperature, the mechanisms depend only on the two renormalized transition rates. We plot the regime where each mechanism holds for a specific inverse temperature, $\beta_b$, and energy level distribution $r$ (see Figure \ref{fig:rot} and \ref{fig:hyd} where $\tilde{a}_{23}= a_{23}/a_{12}$ is represented on the x-axis and $\tilde{a}_{13}=a_{13}/a_{12}$ on the $y$ axis). In this space, a constant $p$ is represented by diagonal lines with a specific angle, a constant $ p\times q$ by vertical lines, and a constant $p\times l$ by diagonal lines with an angle different than those with a constant $p$. At the point $\tilde{a}_{13}=1$ and $\tilde{a}_{23}=e^{-\beta_b(\epsilon_2-\epsilon_1)}$  the lines $p= p\times q=p\times l=0$ cross and the transition matrix has degenerated eigenvalues. Therefore, there is no fast eigenvector at this point, which we call  singular point. Along any straight line starting at the singular point,  the fast eigenvector has the same angle. Therefore, these lines separate regions without the Markovian Mpemba effect, with the direct Markovian Mpemba effect, and with the inverse Markovian Mpemba effect. Moreover, they also separate regions with different mechanisms. The line connecting the origin to the singular point corresponds to parameters that produce a horizontal fast eigenvector, and a vertical line starting at the singular point and going up corresponds to a fast eigenvector with an angle of $\pi/3$ relative to the triangle base. The fast eigenvector angle increases clockwise.

The effect of bath temperature on the Markovian Mpemba effect can be understood by tracking the singular point.  For lower temperatures, the singular point value of $\tilde{a}_{23}$ decreases, reducing the region where the Markovian Mpemba effect occurs (No purple regions in Figures \ref{fig:rot} and \ref{fig:hyd}). Moreover, the angle dividing the direct and inverse Markovian Mpemba regions also changes (see the bottom row in the figures).

Mechanism 1 holds for levels that are equally spaced or increase with energy. This mechanism is the least restrictive of the three. It does not have any constraint on $l$. This is reflected in a larger combination of transition rates that induce it. This mechanism always produces fast eigenvectors with angles between 0 and $\pi/3$. The lowest angles are mainly related to the inverse Markovian Mpemba effect (yellow regions in Figure \ref{fig:rot}) and higher angles to the direct Markovian Mpemba effect (light blue regions in Figure \ref{fig:rot}). This is the only mechanism that holds for equally spaced energy levels.

Mechanism 2  produces a fast eigenvector with an angle between $\pi/3$ and, for high temperatures up to an angle $\arctan[\frac{1+2r}{\sqrt{3}}]$, and for lower temperatures up to a smaller value. This mechanism applies only to increasing energy-level spacing. It is more restrictive than Mechanism 1 due to the restriction on $l$. 

Mechanism 3   has some similarities with Mechanism 2. These include applying only to $r>1$ and being more restrictive than Mechanism 1 due to constraints on $l$. Nevertheless, it also differs fundamentally from Mechanism 2.   It does not exist at high temperatures, but at low temperatures it produces fast eigenvectors with angles ranging from $\pi/3$ to $\arctan[\frac{1+2r}{\sqrt{3}}]$. The importance of this mechanism is that it is the only one acting at low temperatures.  

Finally, Mechanism 4 applies only to 3-level systems with decreasing energy-level spacing, $r<1$. This is the only mechanism that applies to $r<1$. Due to the smaller maximum angle of the quasistatic locus, these systems have a narrower space of transition rates that exhibit the Markovian Mpemba effect (see Figure \ref{fig:hyd}). This is reflected in the fact that for large $\tilde{a}_{13}$, this mechanism does not produce the Markovian Mpemba effect, in contrast with Mechanisms 1 and 2, where the effect is possible for any value of $\hat{a}_{13}$. This mechanism produces fast eigenvectors with angles between 0 and $\arctan[\frac{1+2r}{\sqrt{3}}]<\frac{\pi}{3}$. 

\section{3LS mechanisms and the Markovian Mpemba effect on  N-level systems}

\begin{figure}[htbp]
		\centering
		\includegraphics[width=1\linewidth]{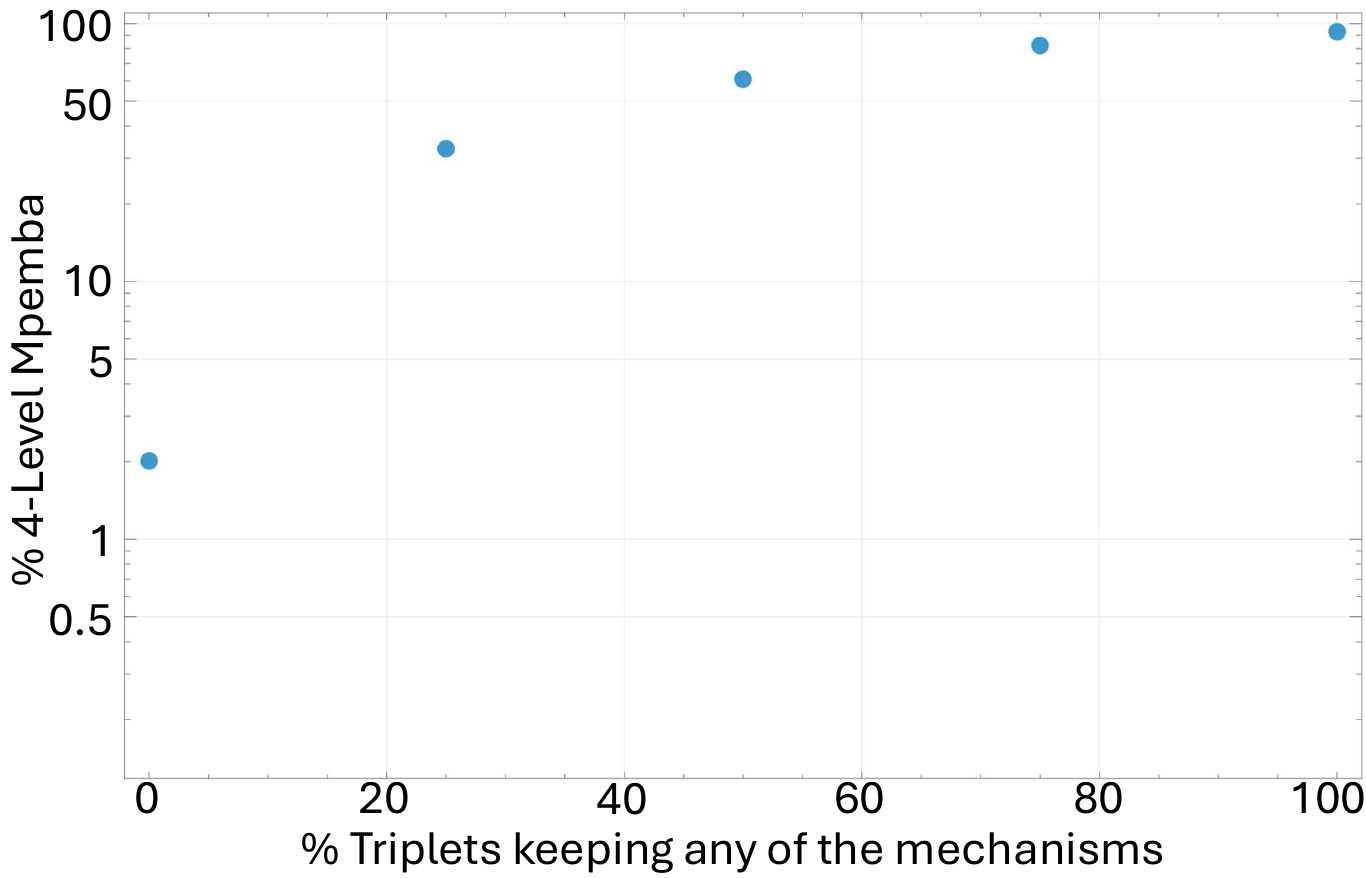}
		\caption{Probability of a 4-level system exhibiting the Markovian Mpemba effect as a function of how many of its triplets exhibit any of the 3LS Markovian Mpemba mechanisms. Notice that even in the case that all the triplets keep one of the 3LS mechanisms, the 4-level system may not exhibit the Markovian Mpemba effect. In this case, the probability is  $\sim93\%$. The convergence of the results was assessed by increasing the number of realizations and confirming that this did not significantly alter the plot. For the transition rates, we considered values from 0.001 to 1 and for $\beta_b \epsilon_i$ from $0$ to $5$. $5\times 10^5$ realizations were used for the whole plot. }	\label{fig:cor}
	\end{figure}

In \cite{avitan2026necessary}, the following necessary conditions for the Markovian Mpemba effect in 3LSs were derived:
	
	\begin{equation}
		\begin{aligned}
			a_{13}C_{31}-a_{23}C_{32} &> a_{12} \left(1-e^{-\beta_b(\epsilon_2-\epsilon_1)}\right)&k\geq1 ; \\
			a_{13}e^{-\beta_b(\epsilon_2-\epsilon_1)}&>a_{23}&k\leq1 ,
            \label{eq:3lsc} 
		\end{aligned} 
	\end{equation}	
  	where $C_{ij}=1+\frac{1}{2}e^{-\beta_b(\epsilon_i-\epsilon_j)}$ and $k\equiv\frac{a_{13} +a_{23}}{a_{12}(1+e^{-\beta_b( \epsilon_2-\epsilon_1)})}$. In the same work, it was shown that these conditions can be used to rule out the Markovian Mpemba effect in N-level systems. This was achieved by dividing the N-level system into all the possible combinations of 3-level systems (called triplets) and applying the 3LS necessary conditions to each triplet. In particular, it was shown that if none of the triplets complies with the necessary 3LS conditions, the N-level system does not experience the Markovian Mpemba effect. Here, we analyze the correlations between the 3LS mechanisms, which are if-and-only-if conditions and the Markovian Mpemba effect in $N$-level systems. To test this, we randomly chose positive values for transition rates, temperatures and energies. Moreover, we impose detailed balance on the transition rates. We then test whether the $N$-level system exhibits the Markovian Mpemba effect and whether its triplets keep any of the 3LS Markovian Mpemba mechanisms. As shown in Figure \ref{fig:cor} for a 4-level system, a higher percentage of triplets exhibiting one of the 3LS mechanisms is correlated with a higher chance of the 4-level system having the Markovian Mpemba effect. But in contrast to the 3LS necessary conditions, we cannot use the 3LS mechanisms to rule out or ensure the Markovian Mpemba effect in an $N$-level system.
    Notice that even in the case that all the triplets keep one of the 3LS Markovian Mpemba mechanisms, this does not ensure that the 4-level system will exhibit the Markovian Mpemba effect. The probability is  $\sim93\%$, which is higher than the probability obtained when all the triplets meet the 3LS necessary conditions \cite{avitan2026necessary}, but still is not 100$\%$. If none of the triplets exhibit any of the 3LS mechanisms,  the probability for the 4-level Markovian Mpemba effect is low, $\sim2\%$, but it is not zero.  In contrast, if none of the triplets keeps the 3LS necessary conditions, the probability for the 4-level Markovian Mpemba effect is zero. 

Nevertheless, our analysis suggests another approach to study the Markovian Mpemba effect in $N$-level systems. This method is based on the singular point. At this point, no fast eigenvector exists because the transition matrix has a single non-zero eigenvalue. 
For a 3LS, consider a circle of small radius centered at the singular point. The transition rates in this circle yield fast eigenvectors at all angles, including those that produce the Markovian Mpemba effect. Therefore, by exploring parameters near the singular point, one can find parameters that produce the Markovian Mpemba effect.

 The singular point is not exclusive to 3LSs. If the
 rates are $a_{ij}=e^{-\beta_b\epsilon_i}$, then 
 all the non-zero eigenvalues of the transition matrix are degenerate. Therefore, there is only a single dynamical speed, and there is no Markovian Mpemba effect. Notice that, at this point, the 3LS necessary conditions, Eq. \eqref{eq:3lsc}, also fail for every triplet: $k=1$ and the l.h.s. and r.h.s of both inequalities have the same value, and therefore the $N$-level can not experience the Markovian Mpemba effect \cite{avitan2026necessary}. At least one triplet should satisfy \eqref{eq:3lsc} to allow the $N$-level system to exhibit the Markovian Mpemba effect. Below, we use this insight to efficiently find parameters for the Markovian Mpemba effect. We achieve this by exploring parameters near the singular point and obtaining triplets that satisfy the Markovian Mpemba necessary conditions.   

There are multiple ways to explore the parameter space around the singular point. Once detailed balance is assumed, an $ N$- level system has $N(N-1)/2$ independent transition rates. Exploring the parameter space of a large system is computationally expensive.  Therefore, we use a simple strategy in which only the forward and backward transition rates of a single pair of states deviate from the value that yields the singular point. We call this protocol single pair of states deviation, or SPSD. For example, if we choose the states $i_0$ and $j_0$, then we take the following transition rates:  $a_{i_0j_0}=e^{-\beta_b\epsilon_{i_0}}(1+\Delta)$,  $a_{j_0i_0}=e^{-\beta_b\epsilon_{j_0}}(1+\Delta)$, and  $a_{ij}=e^{-\beta_b\epsilon_i}$ if at least $i$ or $j$ is different from $i_0$ or $j_0$. The latter is just the singular point transition rate. Notice that the new transition rates still keep detailed balance. As we show below, the SPSD protocol efficiently finds parameters for the Markovian Mpemba effect in large systems. 
 
 The SPSD protocol restores the 3LS necessary conditions for some triplets if applied in the correct ``direction''. To see this consider three energy levels:  $\epsilon_{i_0}<\epsilon_{j_0}$ and $\epsilon_m$. We will consider all the possible orders between the last level and the other two.   If $\epsilon_m$ is the highest or lowest energy level, then only a reduction of $a_{i_0j_0}$, $\Delta<0$,  makes the corresponding triplets fulfill the necessary conditions, Eq. \eqref{eq:3lsc}. In contrast, if $\epsilon_{i_0}<\epsilon_m<\epsilon_{j_0}$, Eq. \eqref{eq:3lsc} holds if and only if $a_{i_0j_0}$ is increased, $\Delta>0$. Notice that by fixing $i_0$ and $j_0$, but not $m$, the arguments above apply simultaneously to $N-2$ triplets.

 To test the SPSD protocol, we consider a specific distribution of energy levels, although part of our results apply universally to any $N$-level system,  as we explain below.  In concrete, we consider a system composed of 20 rotational levels.  For the protocol to be efficient, the forward and backward rate deviations for a randomly chosen pair of states should have a high probability of producing the Markovian Mpemba effect. We determine this probability by counting in how many cases the deviation associated with a single pair of states produces the Markovian Mpemba effect.

\begin{figure}[htbp]
		\centering
		\includegraphics[width=1\linewidth]{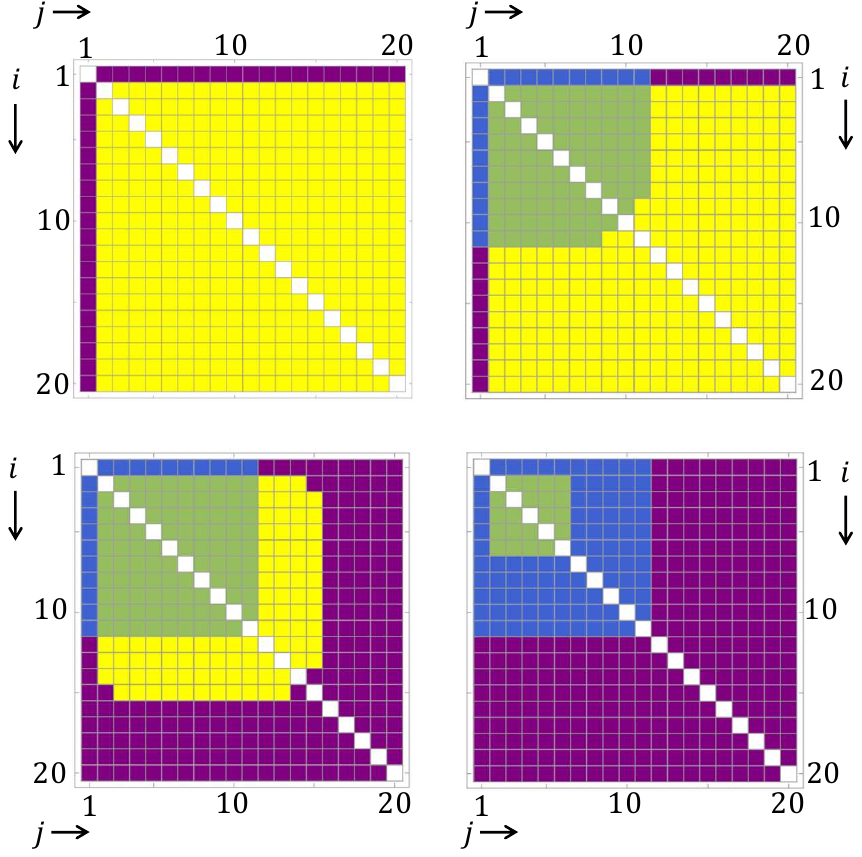}
		\caption{Matrix plots showing the effect of a deviation from the singular point value of the transition rates related to a single pair of states. The color of square $\{i_0,j_0\}$ indicates the effect of only changing the transition rates $a_{i_0j_0}$ and $a_{j_0i_0}$: Blue, yellow, green and purple represent the direct, inverse, direct and inverse, and no Markovian Mpemba effect, respectively. Each matrix represent a different bath temperature: $\beta_b(\epsilon_2-\epsilon_1)=0$, top left; $\beta_b(\epsilon_2-\epsilon_1)=1/10$, top right; $\beta_b(\epsilon_2-\epsilon_1)=1/5$, bottom left and $\beta_b(\epsilon_2-\epsilon_1)=1$, bottom right. For all plots $\Delta=-0.5$, although the plots do not changes as long as  $-1<\Delta<0$.}	\label{fig:mat}
	\end{figure}

 Figure \ref{fig:mat} includes four matrix plots showing when our strategy produces the Markovian Mpemba effect. Each matrix plot is for a different $\beta_b$.  If the square $i_0,j_0$ is purple, then transforming the transition rates $a_{i_0j_0}$ and $a_{j_0i_0}$ does not produce the Markovian Mpemba effect. If the square has another non-white color, it does: yellow for the inverse Markovian Mpemba effect, blue for the direct Markovian Mpemba effect and green for the direct and inverse Markovian Mpemba effect.  
  By dividing the number of off-diagonal squares that produce the Markovian Mpemba effect by the total number of off-diagonal squares, we obtain the probability of success of our protocol when the pair of states is randomly selected. 

\begin{figure}[htbp]
		\centering
		\includegraphics[width=1\linewidth]{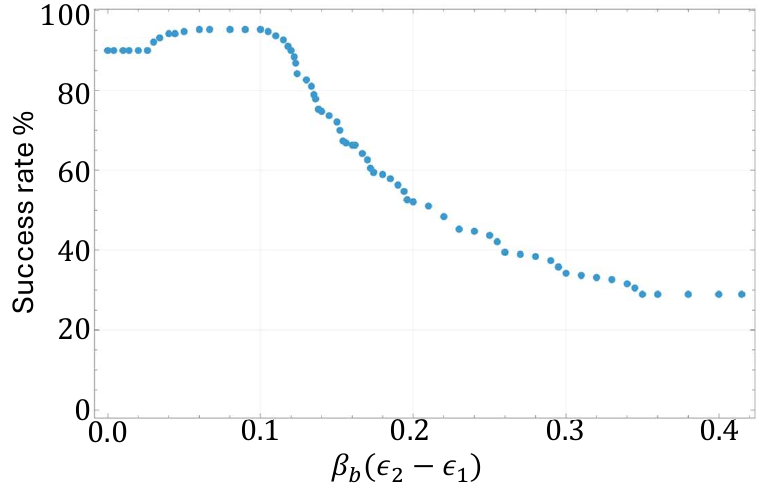}
		\caption{Success rate of the SPSD protocol as a function of the normalized bath inverse temperature. As a system, we consider 20 rotational energy levels, but the high-temperature result applies to any energy level distribution (see the main text).  $\Delta=-0.5$, although the plots do not changes as long as  $-1<\Delta<0$.}	\label{fig:suc}
	\end{figure}

Figure \ref{fig:suc} shows the probability of success as a function of $\beta_b$. We find that in a significant percentage of cases, SPSD produces the Markovian Mpemba effect.   Therefore, we expect this method to be efficient for finding parameters for the Markovian Mpemba effect even for large systems. Moreover, at $\beta_b=0$, the transition matrix is independent of the energy level distribution. Therefore, the high success probability at this temperature, $90\%$, applies to any 20-level system.  The only pairs of transitions that do not yield the Markovian Mpemba effect are those in which either $i_0$ or $j_0$ equals 1.  We always find this pattern regardless of the number of levels. Therefore,  the success rate at $\beta_b=0$ for an $N$-level system is $100\frac{N-2}{N}$, where $N>2$. The larger the system, the higher the probability of success, and as $N\rightarrow\infty$, the success rate approaches $100\%$.

  Analyzing Figure \ref{fig:mat} we can gain insights into the different types of Markovian Mpemba effects.  The direct one is obtained when the selected pair states involve the lowest energy level, and the inverse is obtained for higher energy levels. An intermediate region also exists where both effects are present.   Surprisingly, neither the success probability nor the matrix plot depends on the value of $\Delta$ as long as  $-1<\Delta<0$.
 This effect can be understood from the 3LS cases where the reduction of a rate is equivalent to a translation in a specific direction in the  $\tilde{a}_{23}$- $\tilde{a}_{13}$ plane. If the direction is correct, translations of any size produce the Markovian Mpemba effect as long as the transition rates remain positive. 
 For finite temperatures, a reduction of $a_{12}$ may produce the direct  Markovian Mpemba effect by increasing $\tilde{a}_{23}$ and $\tilde{a}_{13}$. For not too low temperatures,  a reduction of  $a_{23}$ may produce the inverse Markovian Mpemba effect.  
  For $\Delta>0$, our protocol does not find parameters suitable for the Markovian Mpemba effect in the 20 rotational level system, even though it still makes some triplets comply with the 3LS necessary conditions \eqref{eq:3lsc}. This does not imply a contradiction. It just indicates that some triplets are more relevant for the $N$-level Markovian Mpemba effect than others.

  The matrix plots also show that pairs with lower energy, although not the lowest, are the ones that succeed the most. If a random selection is not desired, the best option is to choose $i_0=2$ and $j_0=3$, which achieve the Markovian Mpemba effect at any temperature. Although this result is specifically for rotational levels, as explained above, the $\beta_b=0$ case applies to any energy distribution. 

  Finally, we analyze the protocol success rate as a function of temperature. Although the success rate is already high for $ \beta_b=0$, it initially increases with $\beta_b$. This is related to the fact that pairs of states with either $i_0$ or $j_0$ equal to one seem to yield only the direct Markovian Mpemba effect, which is forbidden at $\beta_b=0$. Once $\beta_b$ increases, the direct Markovian Mpemba effect becomes possible, increasing the success rate.  Eventually, the success rate decreases with $\beta_b$ because pair of states with either large  $i_0$ or $j_0$ stop producing the Markovian Mpemba effect. At large $\beta_b$, the success rate reaches a plateau at $\sim 28\%$. 
  
\section{Conclusions}
We showed that the Markovian Mpemba effect in 3LSs already captures some essential aspects that also apply to $N$-level systems. These include the reduction of the effect's prevalence with bath temperature, the presence of singular points where the dynamics depends on a single time scale, and the existence of parameters near the singular point that produce the effect.  This motivated us to study the effect on 3LSs, where any analysis is simpler than in larger systems.  We find the necessary and sufficient conditions for the Markovian Mpemba effect in 3LSs, understand its physical mechanisms and determine which types of systems they apply to. Moreover, the 3LS study allowed us to identify the singular point and develop an efficient protocol to find parameters that yield the effect for any $N$-level system.

\textit{Acknowledgement.}
We thank  Sergiy Denysov, whose questions inspired this paper.

\onecolumngrid

\setcounter{equation}{0}
\renewcommand{\theequation}{S\arabic{equation}}

\setcounter{figure}{0}
\renewcommand{\thefigure}{S\arabic{figure}}

\setcounter{section}{0}
\renewcommand{\thesection}{S\arabic{section}}

\setcounter{subsection}{0}
\renewcommand{\thesubsection}{\thesection.\arabic{subsection}}
\vspace{1cm}

\newpage
\begin{centering}
{\large \bf Supporting Information}\\
\end{centering}

\section{2D triangular plane}

To map population states and vectors to the 2D triangular plane, we use the following rotation matrix: 
\begin{align}
 R=   \begin{pmatrix}
    -\dfrac{1}{\sqrt{2}} & \dfrac{1}{\sqrt{2}} & 0 \\
    -\dfrac{1}{\sqrt{6}} 
    & -\dfrac{\sqrt{\frac{3}{2}}}{2} + \dfrac{1}{2\sqrt{6}} 
    & \dfrac{\sqrt{\frac{3}{2}}}{2} + \dfrac{1}{2\sqrt{6}} \\
    \dfrac{1}{\sqrt{3}} & \dfrac{1}{\sqrt{3}} & \dfrac{1}{\sqrt{3}}
    \end{pmatrix}. \label{eq:R}
\end{align}

This rotation matrix maps a normalized population vector, $
\left(p_1,p_2,p_3\right)$ to $
\left(\frac{p_2-p_1}{\sqrt{2}},\frac{3p_3-1}{\sqrt{6}},\frac{1}{\sqrt{3}}\right)$, where we assumed that $\sum_{i=1}^3p_i=1$.

The rotated fast eigenvector is 
\begin{align}
    R\bf{V_3}=
    \begin{pmatrix}
        \frac{-a_{12}+a_{13}+a_{23}-a_{21}-\sqrt{(-a_{12}+a_{13}+a_{23}-a_{21})^2-(a_{32}-a_{31})(-2(a_{12}-a_{13}+a_{23})+2a_{21}+a_{31}-a_{32})}}{\sqrt{2}(a_{32}-a_{31})},\sqrt{\frac{3}{2}},0
    \end{pmatrix}.
\end{align}
Moreover, we assume that transition rates satisfy detailed balance, that is $a_{ij}/a_{ji}=e^{-\beta_b(\epsilon_i-\epsilon_j)}$.

\section{Conditions for the Markovian Mpemba effect in  3-level systems}
In this section, we prove that having a fast eigenvector parallel to a tangent of the quasistatic locus is a necessary and sufficient condition for the Markovian Mpemba effect in 3LSs. For simplicity, we consider only the inverse Markovian Mpemba effect. The generalization to the direct Markovian Mpemba effect is straightforward. 

In the systems in question, thermalization is governed by the Pauli rate equation, which is a set of coupled linear differential equations. The initial conditions are determined  by solving the following equation:

\begin{equation}
    \mathbf{P}(\beta)=\mathbf{P}(\beta_b)+b_2(\beta)\mathbf{v_2}+b_3(\beta)\mathbf{v_3} \label{eq:ini}.
\end{equation}

Here, $\mathbf{P}(\beta)$ is a vector whose elements are the populations of the initial thermal state with inverse temperature $\beta$, $\mathbf{P}(\beta_b)$ is the population vector of the final thermal state at the bath inverse temperature $\beta_b$, $b_{2(3)}(\beta)$ is the overlap of the initial state with the slow (fast) eigenvector and $\mathbf{v_{2(3)}}$ are the slow (fast) eigenvectors of the transition matrix. Notice that if $\beta=\beta_b$, then  $b_{2(3)}(\beta)=0$.
   
In \cite{lu2017nonequilibrium} it was proved that a system exhibits the inverse Markovian Mpemba effect if the overlap of the initial state has a non-monotonic behavior with $\beta$ for $\beta>\beta_b$. The non-monotonic behavior requires $\frac{db_2(\beta)}{d\beta}|_{\beta_0>\beta}=0$. In \cite{avitan2026necessary} it was analytically proven that this implies that the fast eigenvector is parallel to the tangent to the quasistatic locus at $\beta_0$. Therefore, a necessary condition for the inverse Markovian Mpemba effect is a fast eigenvector parallel to a vector tangent to the quasistatic locus. Here, we prove that this condition is also sufficient.  

A fast eigenvector parallel to the quasistatic locus requires $\frac{db_2(\beta)}{d\beta}|_{\beta_0>\beta}=0$, but this does not immediately imply a non-monotic behavior of $b_2(\beta)$. This point could alternatively be a saddle point. In this case, the system will not exhibit the Markovian Mpemba effect. Below, we analyze the analytical expression of $b_2(\beta)$ and discard the possibility of $b_2(\beta)$ having a saddle point. In this way, we demonstrate that the parallel fast eigenvector condition is also sufficient for the effect to hold. 

The analytical expression of $b_2(\beta)$ is
\begin{equation}
  b_2(\beta)=\frac{c_1e^{-\beta \epsilon_1}+c_2e^{-\beta \epsilon_2}+c_3e^{-\beta \epsilon_3}}{e^{-\beta \epsilon_1}+e^{-\beta \epsilon_2}+e^{-\beta \epsilon_3}},  
\end{equation}

where $c_i$ are constants that depend on the transition rates, $\beta_b$ and energy levels, but they are independent of $\beta$. 
To test whether $b_2(\beta)$ has a saddle point, we can simplify the expression by dividing $b_2(\beta)$ by $c_1$. This changes the value of $b_2(\beta)$, but does not change the number of saddle points.
Notice that $b_2(\beta)/c_1$ has the form $\frac{f(\beta)}{g(\beta)}$. A saddle point requires a zero first and second derivatives at the same $\beta$. These two conditions imply:
\begin{equation}
    \frac{d^2f(\beta)}{d\beta} g(\beta)=\frac{d^2g(\beta)}{d\beta} f(\beta)
\end{equation}
Solving this equation for the analytical expression of $b_2(\beta)/c_1$ and replacing it in the condition for having a zero first derivative, we get that, for a non-degenerate 3LS, only for $c_1=c_2=c_3=1$ does $b_2(\beta)/c_1$ have a saddle point. Thus, a saddle point requires  $b_2(\beta)/c_1=1$. But  this is inconsistent with the fact that $b_2(\beta_b)=0$ (see Eq. \ref{eq:ini}). Therefore,  $b_2(\beta)$ has no saddle point, and the condition of a fast eigenvector parallel to a vector tangent to the quasistatic locus is necessary and sufficient.  

 \end{document}